\documentclass[pdflatex, 11pt]{article}
\usepackage[english]{babel}
\usepackage{algorithm}
\usepackage{algorithmic}
\usepackage{amsfonts}
\usepackage{amsmath}
\usepackage{amssymb}
\usepackage{bbm}
\usepackage{blindtext}
\usepackage{booktabs}
\usepackage{color}
\usepackage{geometry}
\usepackage{tabularx}
\usepackage{theorem}
\usepackage{times}
\usepackage{url}
\usepackage[pdftex]{graphicx}
\usepackage{bmpsize}
\usepackage{natbib}

\newcommand{\Var}{\text{Var}}

\newcommand{\eg}{\textit{e.g.~}}
\newcommand{\etal}{\textit{et~al}.}

\newcommand{\vp}{\mathbf{p}}
\newcommand{\vq}{\mathbf{q}}

\newcommand{\vu}{\mathbf{u}}

\newcommand{\vy}{\mathbf{y}}

\renewcommand{\d}{\text{d}}

\newcommand{\veta}{\text{\boldmath{$\eta$}}}

\newcommand{\vmu}{\text{\boldmath{$\mu$}}}

\newcommand{\vphi}{\text{\boldmath{$\phi$}}}

\newcommand{\vtheta}{\text{\boldmath{$\theta$}}}
\newcommand{\vtau}{\text{\boldmath{$\tau$}}}

\newcommand{\M}{\mathcal{M}}

\newcommand{\Cov}{\text{Cov}}

\begin{document}

\title{
{\huge
	\textbf{Bayesian approach to Lorenz curve using time series grouped data}
}
}
\author{
	Genya Kobayashi
	\thanks{
		Graduate School of Social Sciences, Chiba University,
		\textit{Email}: \texttt{gkobayashi@chiba-u.jp}
	}\and
	Yuta Yamauchi
	\thanks{
		Graduate School of Economics, The University of Tokyo
	}\and
	Kazuhiko Kakamu
	\thanks{
		Graduate School of Business Administration, Kobe University
	}\and
	Yuki Kawakubo
	\thanks{
		Graduate School of Social Sciences, Chiba University
	}\and
	Shonosuke Sugasawa
	\thanks{
		Center for Spatial Information Science, The University of Tokyo
	}
}
\date{%
	\today
}

\maketitle

\paragraph{Abstract.}
This study is concerned with estimating the inequality measures associated with the underlying hypothetical income distribution from the times series grouped data on the Lorenz curve. 
We adopt the Dirichlet pseudo likelihood approach where the parameters of the Dirichlet likelihood are set to the differences between the Lorenz curve of the hypothetical income distribution for the consecutive income classes and propose a state space model which combines the transformed parameters of the Lorenz curve through a time series structure. 
Furthermore, the information on the sample size in each survey is introduced into the originally nuisance Dirichlet precision parameter to take into account the variability from the sampling. 
From the simulated data and real data on the Japanese monthly income survey, it is confirmed that the proposed model produces more efficient estimates on the inequality measures than the existing models without time series structures. 

\section{Introduction}\label{sec:intro}
The Lorenz curve plays a crucial role in measuring income inequality. 
Given individual household incomes sorted in the ascending order, the Lorenz curve relates the cumulative share of income to the cumulative share of population. 
When individual household income data are available, the Lorenz curve and its associated inequality measures, such as the Gini coefficient, can be accurately estimated by assuming a parametric hypothetical income distribution or using a nonparametric method for the income distribution \citep[see \eg][]{HK02}. 
However, since individual income data may contain sensitive information that may lead to identification of individuals, most national governments and local governments provide the household income data only in the form of grouped data, which contain the summary of income and numbers or proportions of households for predefined income classes. 
Estimating the Lorenz curve based on the grouped data has drawn substantial attention from both theoretical and empirical perspectives. 
See \citet{Ch08} for an overview. 

There are several approaches to the Lorenz curve estimation based on the grouped data in the parametric framework. 
When the grouped level income data, where the numbers of households for the income classes are reported, the popular approach is to assume a  hypothetical income distribution and estimate its parameters from the data. 
\citet{MX95} and \citet{KK03} list the wide range of statistical distributions that have appeared in the context of the income distribution. 
Given a parameter estimate, the Lorenz curves and inequality measures can be analytically or numerically calculated. 
In the likelihood-based estimation, one can use the multinomial likelihood where the cell probabilities  of the multinomial distribution are derived using the distribution function of the assumed income distribution \citep{Mc84}. 
Alternatively, by regarding the thresholds for the income classes as the selected order statistics, the likelihood function can be constructed as the joint density of the order statistics following \citet{DN03} \citep[see][]{NK11, KN18}. 
Using  the theoretical moments of the income distribution, it is also possible to devise a generalised method of moments (GMM) estimator. 
\citet{HGBRC12} and \citet{GH15} proposed the optimal GMM estimators for the parameters of the income distribution from the grouped data.

When the grouped data on the Lorenz curve or proportion of incomes, where the cumulative proportions or proportions of income for the income classes are reported,  are used, a functional form of the Lorenz curve is fit to the data. 
In addition to the Lorenz curves derived from the statistical distributions, such as the lognormal distribution, the researchers have designed many functional forms which are sufficiently flexible and provide analytical forms of the inequality measures. 
See \citet{Sa08} for the list of parametric Lorenz curves. 
It is also possible to model the Lorenz curve semiparametrically as in \citet{RS96} where either the Lorenz curve or the quantile function is expanded using basis functions. 
\citet{CG02} assumed that the expectation of the income share is equal to the difference between the heights of the Lorenz curves evaluated at the two consecutive cumulative proportion of population corresponding to  the thresholds of the income classes and  employed the pseudo Dirichlet likelihood function for the grouped data on the Lorenz curve. 
The parameters of the Lorenz curve is estimated by maximising the Dirichlet likelihood.
\citet{CG05}  considered the Bayesian estimation based on the Dirichlet likelihood and the posterior estimation is carried out using the Markov chain Monte Carlo (MCMC) method. 
More recently, \citet{KK19} proposed the likelihood-free approach to the Lorenz curve based on the approximate Bayesian computation (ABC) that does not rely on the Dirichlet likelihood and can be implemented even when an analytical form of the Lorenz curve is not available. 

The existing literature predominantly focused on estimating the income distribution and its Lorenz curve using a single set of grouped data, typically from a national survey in a chosen year. 
The information contained in the grouped data from a single year is  limited and an estimation result based on such data can be subject to large uncertainty.  
Therefore, by using an appropriate model that uses the grouped data  simultaneously over a certain period of time and borrows information across the time, we can stabilise the estimates and improve the performance. 
Some exceptions are the lognormal state space model considered by \citet{NKO12} and \citet{NK15}.
The observation equation follows the lognormal distribution which is derived from the asymptotic distribution for the linear model based on the selected order statistics. 
The state equation consists of either the AR(1) or random walk process. 
Since their model is specifically designed for the grouped level income data, it cannot be used for the grouped data on the Lorenz curve or proportion of incomes. 
Furthermore, the shape of the lognormal Lorenz curve is known to be very restrictive and hence there is potentiality of other more flexible parametric income distributions and Lorenz curves to be considered. 

The present paper extends the Dirichlet approach proposed by \citet*{CG02, CG05} in such a way that the model can analyse the time series grouped data and the parameters of the Lorenz curve are time varying based on the state space model. 
Specifically, the parameters of the Lorenz curve that appear in the Dirichlet likelihood follow the latent autoregressive processes or random walk processes after an appropriate transformation. 
The precision parameter introduced in the Dirichlet likelihood can depend on the sample size of the survey of each period. 
In the present setting, this precision parameter is a more meaningful parameter that expresses how much we believe overall in the observed data  and can be estimated more stably, because it is now estimated using the data from all periods rather than  from only one sample from the Dirichlet distribution as in the single period setting of \citet{CG02}. 
The numerical examples demonstrate the proposed model manages to reduce a large amount of uncertainty in estimating the Lorenz curve and Gini coefficient. 

The rest of the paper is organised as follows. 
Section~\ref{sec:method}  first introduces the Dirichlet approach and develops the model for the times series grouped data. 
The MCMC method for posterior computation is also described. 
Section~\ref{sec:num} illustrates the proposed method using the simulated data and real data using the monthly grouped data from the Family  Income and Expenditure Survey of Japan. 
Finally, we conclude in Section~\ref{sec:conc}.

\section{Method}\label{sec:method}
\subsection{Lorenz curve estimation based on Dirichlet distribution }
Suppose that the population is divided into the predefined $K$ income classes. 
Let us denote the observed cumulative shares of households for and income for the income classes by $\vp=(p_0=0,p_1,\dots,p_{K-1},p_K=1)'$ and $\vy=(y_0,y_1,\dots,y_{K-1},y_K=1)'$, respectively. 
The cumulative shares of households and income are usually constructed from an income survey on $n$ individual households. 
Let us denote the cumulative distribution function and probability density function of the hypothetical income distribution in the $i$th area by $H(\cdot|\vtheta)$ and $h(\cdot|\vtheta)$, respectively, where $\vtheta=(\theta_{1},\dots,\theta_{d})'$ is the $d$ dimensional parameter vector of the income distribution. 
Then, the Lorenz curve denoted by $L(y|\vtheta)$ is defined by
\[
	L(y|\vtheta) = \frac{1}{\mu} \int_{0}^{y} H^{-1}(z|\vtheta) \d z, \quad y \in [0,1],
\]
where $\mu$ is the mean of the distribution and $H^{-1}(z|\vtheta) = \inf \left\{ x: H(x|\vtheta) \ge z \right\}$.
Once the parameter estimate for $\vtheta$ is obtained, it is possible to calculate the Gini coefficient from
\begin{equation}\label{eqn:gini}
	G= 1 - 2 \int_{0}^{1} L(z|\vtheta) \d z. 
\end{equation}

\citet*{CG02, CG05} assumed that the expectation of the share for an income class is equal to the difference in the values of the Lorenz curve for the two consecutive groups: 
\[
	E[q_{k}] = L(p_{k}|\vtheta)-L(p_{k-1}|\vtheta),\quad k=1,\dots,K,
\]
where $q_{k}=y_{k}-y_{k-1}$ is the income share for the $j$th income class.  
Then,  $\vq=(q_{1},\dots,q_{k})$ is assumed to follow the Dirichlet distribution 
\begin{equation}\label{eqn:dir}
	f(\vq| \vtheta,\lambda) = \Gamma(\lambda ) \prod_{k = 1}^{K}\frac{q_{k}^{\lambda(L(p_{k}|\vtheta) - L(p_{k-1}|\vtheta))- 1}}{\Gamma(\lambda(L(p_{k}|\vtheta) - L(p_{k-1}|\vtheta)))},
\end{equation}
where $\Gamma(\cdot)$ is the gamma function and $\lambda>0$ is the precision parameter. 
The variance and covariance of the income share implied from the Dirichlet distribution are given by
\begin{equation}\label{eqn:var}
	\Var(q_{k}|\vtheta,\lambda)=\frac{E[q_{k}](1 - E[q_{k}])}{\lambda+ 1}, \quad \Cov(q_{k}, q_{l}|\vtheta,\lambda) = -\frac{E[q_{k}]E[q_{l}]}{\lambda + 1},
\end{equation}
A larger value of $\lambda$ suggests that the variation of the income shares implied from the hypothetical Lorenz curve is small. 
Using \eqref{eqn:dir} as the likelihood function, \citet*{CG02, CG05} respectively considered the maximum likelihood estimation and Bayesian estimation assuming a prior distribution for $\vtheta$.

In the setting of \citet*{CG02, CG05}, the data are from, for example, a nation wide income survey for a single year. 
The parameters of the income distribution are estimated based on the Dirichlet likelihood function constructed from a single data point. 
Therefore, the parameter estimates exhibit large variation due to the small sample size. 
Furthermore, the data do not contain information on the Dirichlet precision parameter $\lambda$ and $\lambda$ is seen as a nuisance parameter or tuning parameter. 
Although the value of $\lambda$ can potentially have an impact on the estimates of the parameters and Gini coefficient \citep{KK19}, there exists no clear guideline on the choice of its value when it is fixed nor the choice of its prior distribution when the model is estimated within the Bayesian framework.

\subsection{Model using time series grouped data}
Suppose that the income survey is conducted monthly, quarterly or yearly as is the case in the income survey in Japan (see Section~\ref{sec:real}). 
From a series of the data from the survey we can capture how the income distribution and associated inequality measures change over time. 
In stead of fitting the Dirichlet model \eqref{eqn:dir} independently for each period, we propose to use data from all the available periods and estimate a joint model in order to borrow strength across time and improve the estimation accuracy. 

The notation used in the previous section is indexed by $t=1,\dots,T$ to denote the period. 
The index $t$ is added to \eqref{eqn:dir} as 
\begin{equation}\label{eqn:dirt}
	f(\vq_t| \vtheta_t,\lambda_t) = \Gamma(\lambda_t ) \prod_{k = 1}^{K}\frac{q_{tj}^{\lambda_t(L(p_{tk}|\vtheta_t) - L(p_{t,k-1}|\vtheta_t))- 1}}{\Gamma(\lambda_t(L(p_{tk}|\vtheta_t) - L(p_{t,k-1}|\vtheta_t)))}, \quad t=1,\dots,T,
\end{equation}
where $\vq_t=(q_{t1},\dots,q_{tK})'=(y_{t1}-y_{t0},\dots,y_{tK}-y_{t,K-1})'$ and $\vp_t=(p_{t1},\dots,p_{tK})'$ are the cumulative shares of income and household, $\vtheta_t=(\theta_{t1},\dots,\theta_{td})'$ is the parameters of the income distribution and $\lambda_t$ is the the precision parameter of the Dirichlet distribution at the $t$th period. 
In this paper, the number of income classes and the family of the hypothetical income distribution are assumed to be known and the same for all periods. 
The model \eqref{eqn:dirt} assumes the conditional independence of $\vq_t$ given $\vtheta_t$ and $\lambda_t$. 

The change in the income distribution is captured through the change in the parameters of the income distribution over time. 
This is achieved by modelling a latent process for the appropriately transformed $\vtheta_{t}$ that is to be combined with \eqref{eqn:dirt} to form a state space model. 
Let us define $u_{tj}=h_j(\theta_{tj})$ where $h_j(\cdot)$ is an appropriate link function. 
In the present paper, the parameters on the positive real line are transformed using the log link and those on the unit interval are transformed using the logit link. 
Appendix describes the transformations for constructing the latent processes for the Lorenz curves considered in this paper.

In this paper,  $u_{tj}$ is assumed to independently follow either the AR(1) process
\begin{equation}\label{eqn:ar}
\begin{split}
u_{1j}&\sim N(\mu_j, \tau_j^2/(1-\rho_j^2)),\\
u_{tj}&=\mu_j + \rho_j(u_{t-1,j}-\mu_j) + e_{tj},\quad |\rho_j|<1,\quad e_{tj}\sim N(0,\tau_j^2),\quad t=2,\dots,T,
\end{split}
\end{equation}
or the random walk (RW) process
\begin{equation}\label{eqn:rw}
\begin{split}
u_{1j}&\sim N(0,c\tau_j^2),\\
u_{tj}&=u_{t,j-1} + e_{tj},\quad e_{tj}\sim N(0,\tau_j^2),\quad t=2,\dots,T,
\end{split}
\end{equation}
for $j=1,\dots,d$. 

The precision parameter $\lambda_t$ is modelled proportionally to the sample size $n_t$ of the survey in the $t$th period:
\begin{equation}\label{eqn:lam}
\lambda_t=n_t\exp( \psi),\quad t=1,\dots,T. 
\end{equation}
From \eqref{eqn:var}, the variation of the income shares decreases as the sample size increases. 
While a similar model assumption is found in the context of small area estimation for aggregate data where the similar quantity corresponding to $\exp(-\psi)$ is assumed to be known, the present paper treats $\psi$ as an unknown parameter.  
In the present model, the parameter $\psi$ represents the overall sampling precision  across time and the variation of the income shares varies with the sample size multiplicatively. 
Therefore, as seen in the application the Japanese data, compared to the separate approach of \citet*{CG02,CG05} $\psi$ is a more meaningful parameter and can be estimated more stably as we use the entire time series data to estimate the model.

The parameters of the model \eqref{eqn:dirt} with \eqref{eqn:ar} or \eqref{eqn:rw} and \eqref{eqn:lam} are $(\mu_1,\dots,\mu_d)$, $(\rho_1,\dots,\rho_d)$, $(\tau_1,\dots,\tau_d)$ and $\psi$ and their prior distributions are as follows. 
We assume $\mu_j\sim N(m_j,v_j^2)$ and $\tau_j^2\sim IG(r_j,s_j)$ for the ease of posterior computation for $j=1,\dots,d$. 
The values of the hyperparameters are chosen such that the prior distributions well cover the range of the parameters of the income distribution before transformation found in the existing empirical results (see Section~\ref{sec:num}). 
As we do not have own prior information on the persistency of the transformed parameter of the income distribution, we assume $\rho_j\sim U(-1,1)$ for $j=1,\dots,d$.  
Finally, as we have little prior information on $\psi$, we assume $\psi\sim N(m_\psi,v^2_\psi)$ with the hyperparameters such that the prior distribution covers a wide range of values of $\psi$. 

Finally, the joint distribution of the data $\vq=(\vq_1,\dots,\vq_T)$, the latent variables $\vu=(\vu_1,\dots,\vu_{T})$ with $\vu_t=(u_{t1},\dots,u_{td})'=(h^{-1}(\theta_{t1}),\dots,h^{-1}(\theta_{td}))'$ and parameters $\veta_j=(\mu_j,\rho_j,\tau^2_j)'$ and $\psi$ under the AR(1) specification \eqref{eqn:ar} is given by
\begin{equation*}
\pi(\vq,\vu,\vmu,\vphi,\vtau^2,\psi)\propto \left[\prod_{t=1}^Tf(\vq_t|\vu_t,\psi) \left\{\prod_{j=1}^d\pi(u_{tj}|u_{t-1,j},\veta_j )\right\}\right] \pi(\psi)\prod_{j=1}^d\left\{ \pi(\mu_j)\pi(\rho_j)\rho(\tau^2_j)\right\},
\end{equation*}
where $\pi(u_{tj}|u_{t-1,j},\veta_j )$ is the density function associated with the model \eqref{eqn:ar} with $\pi(u_{1j}|u_{0,j},\veta_j )=\pi(u_{1j}|\veta_j )$ and $\pi(\mu_j)$, $\pi(\rho_j)$, $\pi(\tau^2_j)$ and $\pi(\psi)$ are the prior densities. 
The joint distribution for the RW specification \eqref{eqn:rw} is given by
\begin{equation*}
\pi(\vq,\vu,\vtau^2,\psi)\propto \left[\prod_{t=1}^Tf(\vq_t|\vu_t,\psi) \left\{\prod_{j=1}^d\pi(u_{tj}|u_{t-1,j},\tau_j^2 )\right\}\right] \pi(\psi)\prod_{j=1}^d \left\{\pi(\tau^2_j)\right\},
\end{equation*}
with $\pi(u_{1j}|u_{0,j},\tau_j^2)=\pi(u_{1j}|\tau_j^2)$.

\subsection{Posterior computation}
The proposed model is estimated using the MCMC method based on the simple Gibbs sampler. 
Here the sampling algorithm for the AR(1) specification is described. 
The sampling algorithm for the RW specification can be easily obtained after a few modification. 
The parameters $(\mu_1,\dots,\mu_d)$, $(\rho_1,\dots,\rho_d)$, $(\tau_1,\dots,\tau_d)$ and $\psi$ and latent variables $\vu_t,\ t=1,\dots,T$ are alternately sampled from their respective full conditional distributions as follows. 

\begin{enumerate}
\item
We sample $\vu_t$ from the full conditional distribution proportional to 
\begin{equation*}
\pi(\vu_t|\text{Rest})\propto\left\{
\begin{split}
& f(\vq_1|\vu_1,\psi)\prod_{j=1}^d\left\{\pi(u_{1j}|\veta_j)\right\},\quad t=1\\
& f(\vq_T|\vu_T,\psi)\prod_{j=1}^d\left\{\pi(u_{Tj}|u_{T-1,j},\veta_j)\right\},\quad t=T\\
 &f(\vq_t|\vu_t,\psi)\prod_{j=1}^d\left\{\pi(u_{tj}|u_{t-1,j},\veta_j )\pi(u_{t+1,j}|u_{t,j},\veta_j )\right\}, \quad \text{otherwise}. 
 \end{split}
 \right.
\end{equation*}
It is not in the form of the density function of a standard distribution due to the complex form of the Dirichlet distribution in $f(\vq_t|\vu_t,\psi)$. 
The accept-reject Metropolis-Hastings (ARMH) algorithm which uses a normal approximation of $\pi(\vu_t|\text{Rest})$ around the mode is adopted. 
As necessary,  the random walk Metropolis-Hastings (MH)  algorithm may be mixed with probability 0.05 during the burn-in period in order to escape from badly chosen initial values.

\item
From $j=1,\dots,d$, the full conditional distribution of $\mu_j$ is $N(\hat{m}_{\mu_j},\hat{v}^2_{\mu_j})$ where
\begin{equation*}
\hat{m}_{\mu_j}=\hat{v}^2_{\mu_j}\left\{\frac{1-\rho_j^2}{\tau_j} u_{1j}+\frac{1-\rho_j}{\tau^2_j}\sum_{t=2}^T(u_{tj}-\rho_ju_{t-1,j}) + \frac{m_j}{v_j^2}\right\},\quad 
\hat{v}_{\mu_j}^2=\left\{\frac{(T-1)(1-\rho^2_j)}{\tau_j^2} + \frac{1-\rho^2_j}{\tau_j^2} + \frac{1}{v_j^2}\right\}. 
\end{equation*}

\item
The full conditional distribution of $\rho_j$ is proportional to 
\begin{equation*}
\begin{split}
\pi(\rho_j|\text{Rest})&\propto \sqrt{1-\rho_j^2} \exp\left\{-\frac{(1-\rho_j^2)(h_{1j}-\mu_j)^2}{2\tau_j^2}\right\}\exp\left\{-\frac{1}{2\tau^2_j} \sum_{t=2}^T ( u_{tj}-\mu_j-\rho_j(u_{t-1,j}-\mu_j))^2\right\}\pi(\rho_j)\\
&\propto \sqrt{1-\rho_j^2}\exp\left\{-\frac{(\rho_j-\hat{m}_{\rho_j})^2}{2\hat{v}_{\rho_j}^2}\right\}\pi(\rho_j).
\end{split}
\end{equation*} 
The independence MH algorithm is used to sample $\rho_j$ with the proposal distribution $N(\hat{m}_{\rho_j},\hat{v}_{\rho_j}^2)$ where
\begin{equation*}
\hat{m}_{\rho_j}=\hat{v}_{\rho_j}^2\left\{\frac{1}{\tau_j^2}\sum_{t=2}^T (u_{tj}-\mu_j)(u_{t-1,j}-\mu_j) \right\},\quad
\hat{v}_{\rho_j}^2=\frac{\tau_j^2}{\sum_{t=2}^{T-1} (u_{tj}-\mu_j)^2} .
\end{equation*}
With the current state $\rho_j$ and proposal $\rho_j^*$, the acceptance probability is given by
\begin{equation*}
\min\left\{1,\frac{\sqrt{1-\rho_j^{*2}}\pi(\rho_j^*)}{\sqrt{1-\rho_j^2}\pi(\rho_j)} \right\}.
\end{equation*}

\item
For $j=1,\dots,d$, the full conditional distribution of $\tau_{tj}$ is given by $IG(\hat{r}_j, \hat{s}_j)$ where
\begin{equation*}
\hat{r}_j=r_j + \frac{T}{2},\quad 
\hat{s}_j = s_j + \frac{1}{2}\left[ (1-\rho_j^2)(u_{1j}-\mu_j)^2 + \sum_{t=2}^T (u_{tj}-\mu_j-\rho_j(u_{t-1,j}-\mu_j))^2 \right].
\end{equation*}

\item
The full conditional distribution of $\psi$ is proportional to
\begin{equation*}
\pi(\psi|\text{Rest})\propto \left\{\prod_{t=1}^T f(\vq_t|\vu_t,\psi)\right\}\pi(\psi). 
\end{equation*}
Again, since this is not in a standard form, we use the random walk MH algorithm  with a normal proposal distribution to sample $\psi$.  

\end{enumerate}

Combining different parametric income families with different models for the latent process leads us to fitting multiple models to the data. 
The models are compared based on a criterion based on the posterior predictive loss proposed by \citet{GG98}.
The criterion favours the model $\M$ which minimises
\begin{equation*}
\text{PPL}_r(\M)=\sum_{t=1}^T\sum_{k=1}^K V^\M_{tk}+\frac{r}{r+1}\sum_{t=1}^T\sum_{k=1}^K(q_{tk}-E^\M_{tk})^2
\end{equation*}
where $E^\M_{tk}$ and $V^\M_{tk}$ are the mean and variance of the posterior predictive distribution of $q_{tk}$ under the model $\M$. 
We estimate $E^\M_{tk}$ and $V^\M_{tk}$ based on the 10000 MCMC draws obtained after the 2000 draws of the burn-in period. 
The first term penalises the model for complexity and the second term measures the goodness of fit with the weight $r$.

\section{Numerical Examples}\label{sec:num}
\subsection{Simulation study}\label{sec:sim}
The proposed method is first illustrated using the simulated datasets. 
This study considers the Singh-Maddala distribution, denoted by $SM(\alpha,\beta,\gamma)$, as the hypothetical income distribution. 
This is a widely used three parameter family and is known to fit to income data well. 
Since the Lorenz curve of the Singh-Maddala distribution does not depend on the scale parameter $\beta>0$, only are the latent processes for the shape parameters $\alpha,\gamma>0$ modelled. 
The latent variables are $u_{t1}=\log\alpha_t$ and $u_{t2}=\log\gamma_t$ for $t=1,\dots,T$. 
\ref{sec:inc} provides the probability density function and associated inequality measures of the income distributions used in this paper. 

The simulated data are generated as follows. 
\begin{enumerate}
\item
The latent variables $\vu_t$, are generated based on the AR(1) process given by \eqref{eqn:ar}. 
\item
The sample size $n_t$ is selected randomly from $\{5000,6000,\dots,15000\}$. 
\item
Given $\vu_t$ and $n_t$,  the household incomes, denoted by $x_{t1},\dots,x_{tn_t}$,  are generated from  $SM(\alpha_t,1,\gamma_t)$. 
\item
The household incomes are sorted in the ascending order $x_{t(1)},\dots,x_{t(n)}$ and divided into equally sized $K$ income classes, then   the income shares are computed from $q_{tk}=\sum_{i=m_{k-1}+1}^{m_{kt}} x_{t(i)}/\sum_{j=1}^{n_t} x_{tj}$ for $k=1,\dots,K$ where $m_0=0$, $m_{kt}=kn_t/K$ and $m_{Kt}=n_t$.  
\end{enumerate}

In this simulation study, we set $K=5$ so that $\vp=(0,0.2,0.4,0.6,0.8,1)'$ as in the income survey data in Japan (Section~\ref{sec:real}) and $T=500$. 
We set $\eta_1=(1.25,0.8,0.015)'$ and $\eta_2=(0.4,0.8,0.02)'$. 
The prior distributions are given by $\mu_j\sim N(0,1)$, $\tau^2_j\sim IG(3,0.1), \ j=1,\dots,d$ and $\psi\sim N(0,100)$ such that the prior distributions well cover the range of the parameter values often found in the literature. 
The MCMC algorithm are run for 10000 iterations after a burn-in period of 2000 iterations. 

Table~\ref{tab:sim1} presents the posterior means, 95\% credible intervals and inefficiency factors of the parameters. 
The table shows that the posterior distributions are concentrated around the true values. 
The inefficiency factor is defined as a ratio of the numerical variance of the sample mean of the Markov chain to the variance of the independent draws \citep{Ch01}. 
The inefficiency factors shown in the table are reasonably small indicating the efficiency of the sampling algorithm. 

Using the proposed method, the Gini coefficient and values of the Lorenz curve for $p=0.2,0.4,0.6,0.8$ are estimated based on the posterior sample. 
The performance of the proposed method is compared with the two alternative approaches. 
The first alternative is a crude descriptive statistics which estimates the Gini coefficient from based on the trapezoids for each period given by
$
\sum_{k=0}^K(y_{tk}+y_{t,k-1})(p_k-p_{k-1})/2, \ t=1,\dots,T.
$
The second alternative is the method of \citet{CG05}, which estimates the parameters $(\alpha_t,\gamma_t,\log{\lambda}_t)$ separately for $t=1,\dots,T$ based on the Dirichlet likelihood assuming the independent $N(0,1)$ priors for $\log\alpha_t$ and $\log\gamma_t$, and $N(0,100)$ prior for $\log\lambda_t$. 
In this approach, no information on the sample size is reflected to the precision parameter $\lambda_t$. 
For each period, the random walk MH algorithm which jointly samples $(\alpha_t,\gamma_t,\lambda_t)'$ is run for 10000 iterations after a burn-in period of 2000 iterations. 
Given the posterior samples on the parameters, the Gini coefficients and Lorenz curves are estimated. 
The performance of the three approaches are compared based on the relative bias, defined by $(\hat{G}_t-G_t)/G_t$ where $\hat{G}_t$ is the posterior mean under a method and $G_t$ is the true value of the Gini coefficient, and lengths of the credible intervals over time. 
We also considered a more diffuse prior distributions $N(-0.5,2)$ for $\mu_j$, $\log\alpha_t$ and $\log\gamma_t$. 
They have the same prior means as the default prior  but the inflated prior variances.

\begin{figure}[H]
\centering
\includegraphics[scale=0.55]{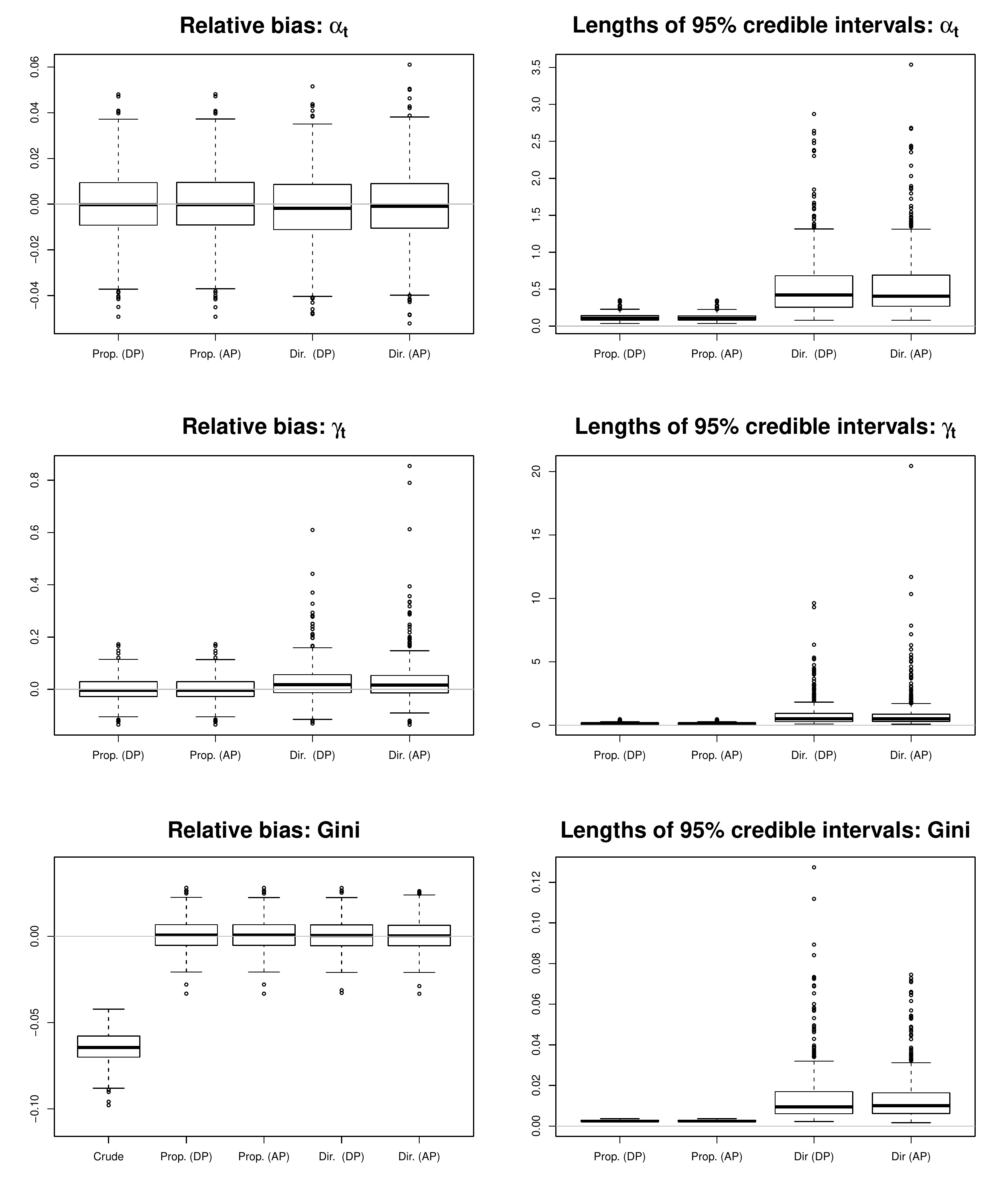}
\caption{The boxplots of the relative bias (left) and lengths of the credible intervals (right) for $\alpha_t$, $\gamma_t$ and Gini coefficients under the proposed approach (Prop.), separate Dirichlet approach (Dir.) and crude descriptive statistics (Crude) over $T=500$ periods for the simulated data with the default $N(0,1)$ prior (DP) and alternative $N(-0.5,2)$ prior (AP) for $\mu_j$ for the proposed approach, and  $\log\alpha_t$ and $\log\gamma_t$ for the separate approach }
\label{fig:sim1}
\end{figure}

Figure~\ref{fig:sim1} presents the boxplots of the relative bias and lengths of the 95\% credible intervals for $\alpha_t$, $\gamma_t$ and the Gini coefficients under the proposed approach, separate Dirichlet approach of Chotikapanich and Griffiths~(2005) and crude descriptive statistics over $T=500$ periods.
The relative bias for $\alpha_t$ under the proposed and separate approaches appear to be comparable and the estimates do not seem to depend on the prior specification. 
This is also the case for the Gini coefficient.  
However, the figure shows that estimating the parameters separately can occasionally incur large bias in the estimates of $\gamma_t$ and  the amount of bias appears to depend the prior specification. 
The figure shows that the 95\% credible intervals for the parameters and Gini coefficients under the proposed approach are immeasurably narrower than those under the separate Dirichlet approach, indicating large reduction in the uncertainty on those quantities by  borrowing strength across time through the state space model.  
The wide credible intervals under the separate Dirichlet approach is resulted from the large uncertainty in the posterior distributions because the data from a single period contains little information about the parameters. 
It is also seen that the lengths of the credible intervals under the separate approach tend to become longer as the prior variances increases. 
On the other hand, the proposed approach appears to be robust with respect to the prior setting. 

\begin{figure}[H]
\centering
\includegraphics[scale=0.53]{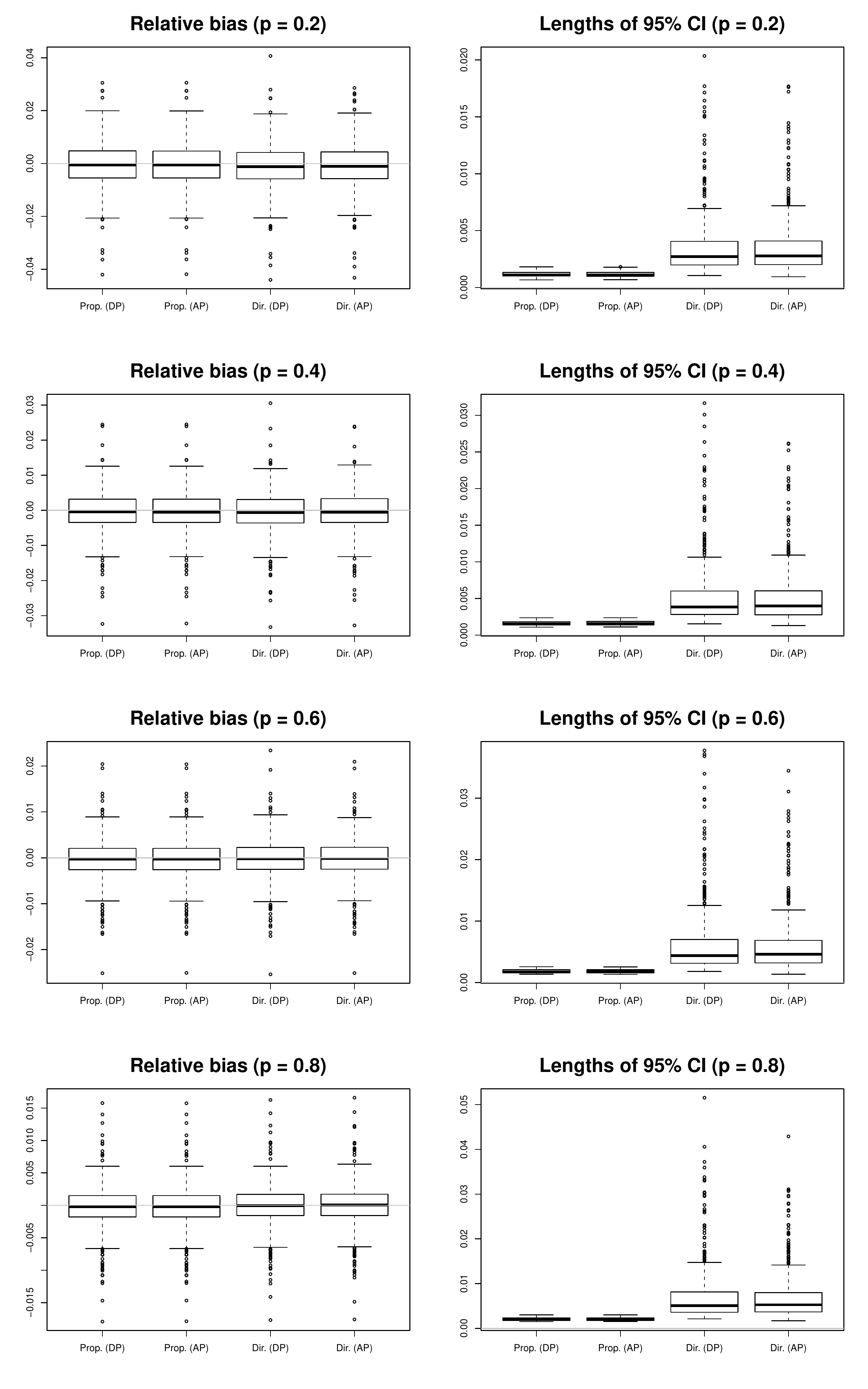}
\caption{The boxplots of the relative bias (left) and lengths of the credible intervals (right) for the Lorenz curve with $p=0.2,0.4,0.6,0.8$ under the proposed approach (Prop.) and separate Dirichlet approach (Dir.) over $T=500$ periods for the simulated data with the default $N(0,1)$ prior (DP) and alternative $N(-0.5,2)$ prior (AP) for $\mu_j$ for the proposed approach, and $\log\alpha_t$ and $\log\gamma_t$ for the separate approach}
\label{fig:sim2}
\end{figure}

The same observation holds for the estimates of the Lorenz curve. 
Figure~\ref{fig:sim2} presents the relative bias and lengths of the 95\% credible intervals for the Lorenz curves at $p=0.2,0.4,0.6, 0.8$ under the proposed and separate Dirichlet approaches. 
The left panels of the figure show that the relative bias under the proposed and separate approaches are comparable. 
However, the right panels show that the credible intervals under the proposed approach are much narrower than those under the separate approach and the prior specification under the separate approach has an impact on the lengths of the intervals.


\begin{table}[H]
\centering
\caption{The posterior means, 95\% credible intervals (CI) and inefficiency factors for the simulated data}
\label{tab:sim1}
\begin{tabular}{ccrrrr}\hline\hline
\multicolumn{1}{c}{Parameter}& True & \multicolumn{1}{c}{Mean} & \multicolumn{2}{c}{95\% CI} & \multicolumn{1}{c}{IF}\\\hline
$\mu_1$     & 1.25    &  1.247  &  (   1.189, &   1.306)  &  0.988 \\
$\mu_2$     &  0.4    &  0.377  &  (   0.349, &   0.405)  &  1.058 \\
$\rho_1$    &  0.8    &  0.814  &  (   0.762, &   0.866)  &  0.896 \\
$\rho_2$    &  0.5    &  0.537  &  (   0.460, &   0.613)  &  1.279 \\
$\tau_1$    &  0.015  &  0.014  &  (   0.012, &   0.016)  &  0.709 \\
$\tau_2$    &  0.02   &  0.022  &  (   0.019, &   0.025)  &  1.293 \\
$\psi$      &  -----  &  4.428  &  (   4.340, &   4.514)  &  5.349 \\
\hline
\end{tabular}
\end{table}

\subsection{Application to Japanese income survey data}\label{sec:real}
\begin{figure}[H]
\centering
\includegraphics[width=\textwidth]{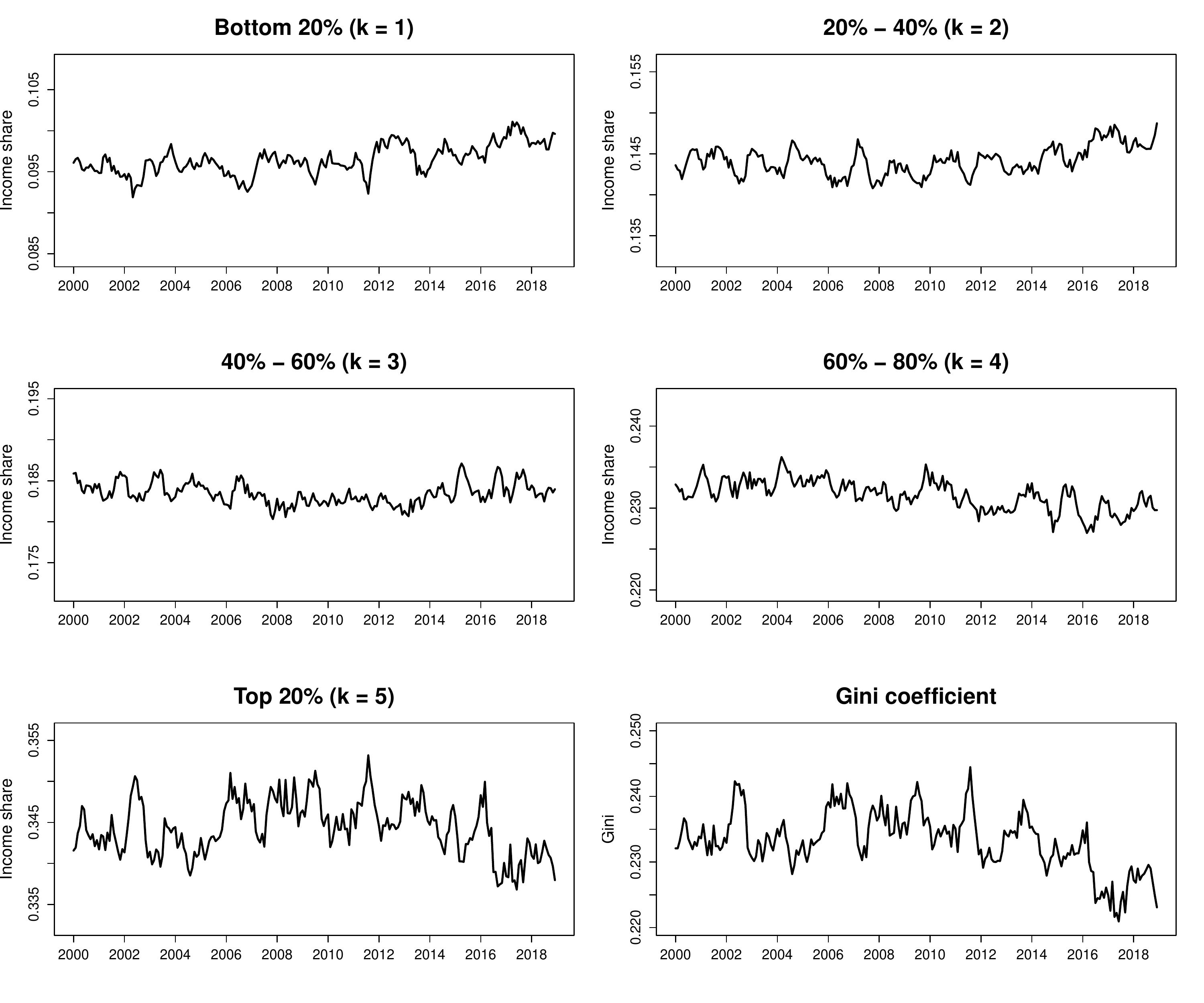}
\caption{Time series plot of the income shares of the income classes $k=1,\dots,5$ and crude descriptive statistics of the Gini coefficient for the FIES data}
\label{fig:real1}
\end{figure}

Now the proposed model is demonstrated using the monthly income share data from the Family Income and Expenditure Survey (FIES) prepared by Ministry of Internal Affairs and Communications of Japan. 
Our dataset contains the income shares of the $K=5$ equally sized income classes of the $n_t=n=10000$ working households which has been adjusted to the population size between  January 2000 and December 2018 ($T=228$). 
The dataset is available from https://www.e-stat.go.jp. 
Figure~\ref{fig:real1} presents the time series plot of the income shares of  the five income classes of our dataset and the descriptive statistics of the Gini coefficient. 
The figure shows that the share of the fifth (highest income) class exhibits some large variation around 0.35. 
The shares of the other four classes appear to be rather stable over time with smaller variation. 
Therefore, the figure suggests fitting an income model that well captures the behaviour of the upper tail of the distribution is important to obtain an accurate insight about the dynamics of the income distribution and inequality structure. 
We consider the three Lorenz curves derived from the  hypothetical income distribution, lognormal (LN), Singh-Maddala (SM) and Dagum (DA) distributions and three parametric functional Lorenz curves proposed by  \citet{OFLG91} (OR) and \citet{RGKO80} (RA) with two parameters and \citet{Ka80} (KA) with three parameters. 
Each Lorenz curve is incorporated the proposed framework where the parameters follow the latent AR(1) or RW. 
The transformations for constructing the latent processes for those Lorenz curves are described in Appendix. 
The separate Dirichlet approach (DIR) is fitted as well. 
In total eighteen models are fitted to the data. 
Hereafter, the models are denoted as, for example, LNAR for the LN model with AR(1) and KADIR for the KA model using the separate approach, as necessary. 
The same default prior distribution as in the simulation study. 
For the RW models, we set $c=10^{5}$. 
The models are estimated based on the 10000 MCMC draws after the 2000 draws of the burn-in period.

Table~\ref{tab:real1} presents the log PPL for the eighteen models. 
For both $r=1$ and $\infty$, KARW for the latent process resulted in the smallest PPL followed by KAAR version with the very marginal differences. 
In \citet{CG02} the KA Lorenz curve was also found to be the best fitting model. 
The SM models follow the KA models. 
In the proposed approach, the DA models are the least supported by the data based on PPL. 
The Dagum distribution has more control on the shape in the left tail than the right tail \citep{Ka16}. 
This result suggests that the right tail of the income distribution for this particular dataset is relatively more important than the left tail, while the Dagum distribution is known to fit well to income data of many other countries (Kl08). 
For all the other Lorenz curves, the RW specification resulted in smaller PPL than the AR. 
In the following, we only focus on the result under the RW models. 
The PPL under DIR are larger than the proposed approaches due to the large uncertainty from the single period data. 
Among the models with the separate approach, LN resulted in the smallest PPL as it has only one parameter to estimate in the Lorenz curve per period.

\begin{figure}[H]
\centering
\includegraphics[width=\textwidth]{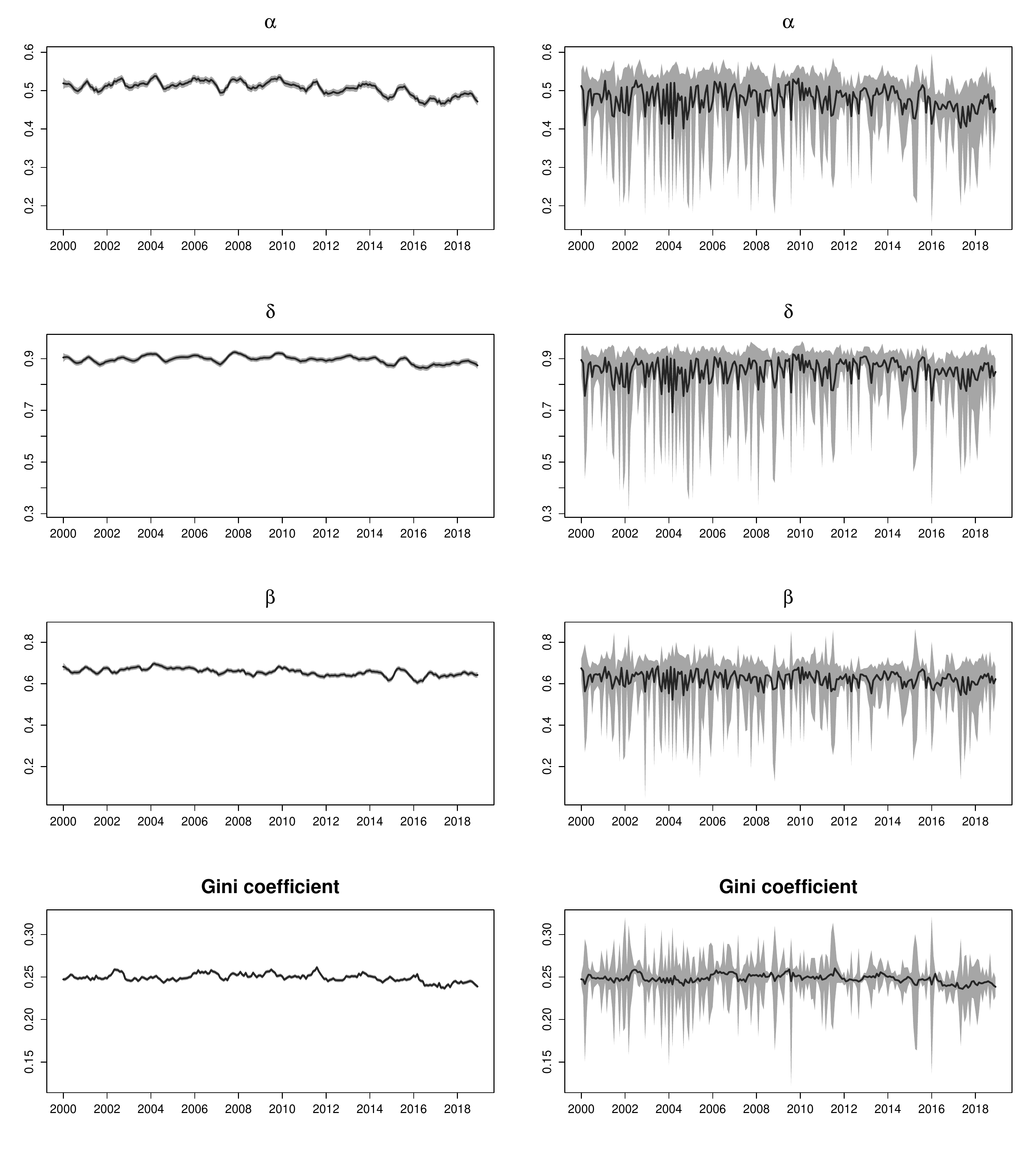}
\caption{The posterior means (solid lines) and 95\% credible intervals (shaded areas) of the parameters and Gini coefficients for KARW (left) and KADIR (right)}
\label{fig:real2}
\end{figure}

Figure~\ref{fig:real2} presents the posterior means and 95\% credible intervals of the parameters and Gini coefficients under KARW and KADIR.
As illustrated in the simulation study, the posterior distributions under the separate approach are wide spread across the parameter spaces. 
Contrary, the proposed approach produced the more concentrated posterior distributions with the smoother traces of the posterior means and hence would provide more precise insight about the dynamics of the quantities of interest.

\begin{figure}[H]
\centering
\includegraphics[width=\textwidth]{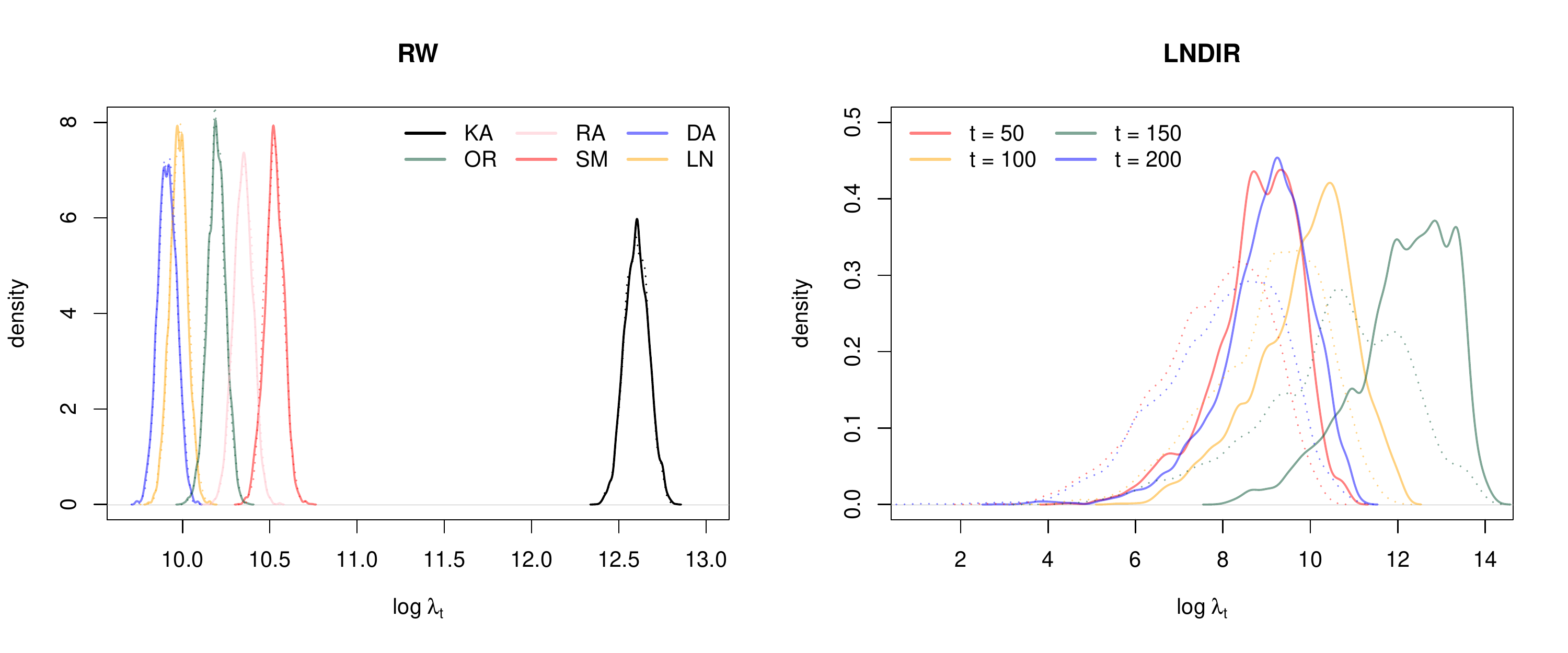}
\caption{The posterior distributions of $\log \lambda_t$ under the RW models (left) and LNDIR  (right) for $t=50, 100, 150, 200$ with the priors $\psi,\log\lambda_t\sim N(0,100)$ (solid) and $N(0,10)$ (dotted)}

\label{fig:real3}
\end{figure}

Figure~\ref{fig:real3} presents the posterior distributions of $\log\lambda_t$ obtained from the proposed approach with RW and  LNDIR, which is the best supported model under the separate approach for $t=50,100,150$ and $200$ under the two prior settings for $\psi$ and $\log \lambda_t$: $N(0,100)$ and $N(0,10)$. 
In the left panel of the figure, it is seen that the locations of the posterior distributions correspond to the order of the model supported by the data shown in Table~\ref{tab:real1}.
Therefore, under the proposed approach, the precision parameter $\psi$ is indicative of how much we believe overall in the observed data. 
Also the posterior distributions seem to robust with respect to the prior specification. 
Contrary, the posterior distributions under the separate approach are wide spread and exhibit prior sensitivity, as also demonstrated by \citet{KK19}.

\begin{figure}[H]
\centering
\includegraphics[width=\textwidth]{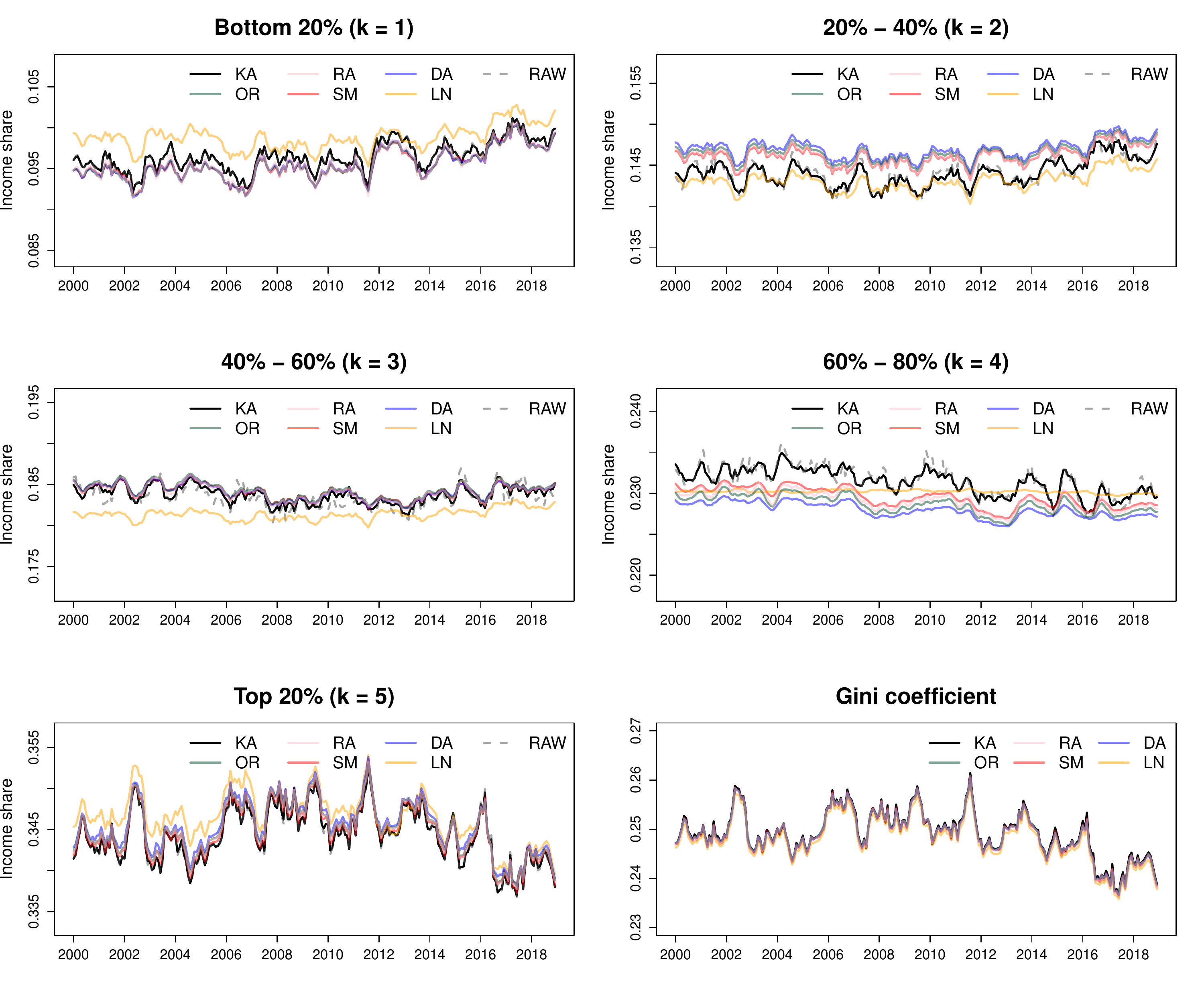}
\caption{The posterior means of the income shares for $k=1,\dots,5$ and  Gini coefficients under the RW models overlaid on the FIES data (RAW)}
\label{fig:real4}
\end{figure}

Figure~\ref{fig:real4} presents the posterior means of the income shares for the five income classes and Gini coefficients under the proposed approach.  
It is seen that KA, which is the most supported by the data based on PPL, traces the observed income shares most closely. 
In the figure, KA, SM, DA, OR and RA seem to agree for $k=1, 3$ and $5$, but some disagreement between the first and latter models are observed for $k=2$ and $4$. 
For $k=1,3$ and $4$, LN resulted in the somewhat peculiar patterns especially in the case of $k=3$ where the income share exhibits little variation over time. 
This could be to the restrictive shape of the lognormal Lorenz curve which is controlled by the single parameter. 
Regardless of some disagreement in the estimates among the six models, the posterior means of the Gini coefficients resulted in the similar patterns, though the estimates under the lognormal model appear to be constantly slightly smaller than the rest. 
The deviations in the estimates of the income share, for example, under SM, DA, OR, RA  from the observed share for $k=2$ and $3$ compensate each other such that they still agree in the Gini coefficient, which is a summary measure of the Lorenz curve. 
The figure also shows that the dynamics of the top 20\% share is closely linked with that of the Gini coefficient. 
It appears that the top 20\% share and Gini coefficient is declining from the middle of the sample period  towards the end with the relative increase in the shares of the bottom two classes.


\begin{table}[H]
\centering
\caption{The log posterior predictive losses with the different weights ($r=1$ and $\infty$) under the proposed AR and RW models and existing DIR model for the FIES data}
\label{tab:real1}
\begin{tabular}{ccrrrrrr}\hline\hline
Weight  & Model & \multicolumn{1}{c}{LN}& \multicolumn{1}{c}{SM}& \multicolumn{1}{c}{DA} & \multicolumn{1}{c}{KA} &  \multicolumn{1}{c}{OR} & \multicolumn{1}{c}{RA} \\\hline
$r=1$     & AR     & -4.348 & -4.863 & -4.240 & -6.868 & -4.522 & -4.665\\
          & RW        & -4.364 & -4.871 & -4.262 & -6.873 & -4.551 & -4.687\\
          & DIR & -3.209 & -2.983 & -2.341 & -1.859 & -2.551 & -2.765\\
$r=\infty$& AR     & -4.144 & -4.684 & -4.040 & -6.744 & -4.329 & -4.497\\
          & RW        & -4.157 & -4.691 & -4.060 & -6.747 & -4.354 & -4.516\\
          & DIR & -3.139 & -2.955 & -2.300 & -1.836 & -2.522 & -2.733\\
\hline
\end{tabular}
\end{table}

\section{Conclusion}\label{sec:conc}
This paper has developed the model for the Lorenz curve based on the time series grouped data where the likelihood part is based on the Dirichlet distribution and the parameters of the Lorenz curve follow the latent process constituting a state space model. 
The numerical examples using the simulated data and real data of Japan have demonstrated that the parameters can be estimated more stably and the posterior distributions obtained from the proposed approach have much smaller uncertainty than separately estimating the parameters based on the Dirichlet distribution using the data only from a single period. 
The application to the income survey data of Japan found that the three-parameter parametric Lorenz curve of \citet{Ka80} provided the best fit to the data followed by the Singh-Maddala Lorenz curve. 
It was also seen that in the sample period between 2000 and 2018 the Gini coefficient exhibit a downward trend after 2008 with some occasional spikes.

Some limitations of the present paper is as follows. 
Although the parametric functional forms of the Lorenz curve fit to data very well, these Lorenz curves cannot provide very detailed insights about the income distribution such as its shape. 
An advantage of the Lorenz curve derived from the statistical distribution is that the shape of the income distribution may be grasped. 
However, since the Lorenz curve is location-free, we cannot obtain the complete picture of the income distribution of interest. 
In order to address this issue,  we could consider extending the GMM approach of \citet{HGBRC12} and \citet{GH15} by utilising an additional information on the location of the income distribution. 
Furthermore, although we could obtain some insights about the dynamics of the income shares and Gini coefficient from the application, one may wish to smooth out the estimates even more for better interpretability using, \eg a regime- or shrinkage-based  modelling. 
These are left for the future studies.

\paragraph{Acknowledgments.}
This work is partially supported by JSPS KAKENHI (\#17J04715, \#18K12754,  \#18K12757,  \#19K13667). 
The computational results were obtained by using Ox version 6.21 \citep{D07}.

\appendix
\def\thesection{Appendix}
\section{Parametric Lorenz curves}\label{sec:inc}
\subsection*{Lognormal distribution}
The probability density  function of the lognormal distribution denoted by $LN(\mu,\sigma^2)$  is given by
\begin{equation*}
h_{LN}(x|\mu,\sigma^2)=\frac{1}{x\sqrt{2\pi\sigma^2}}\exp\left\{-\frac{(\log x - \mu)^2}{2\sigma^2}\right\},\quad x>0,
\end{equation*}
where $\mu$ is the location parameter and $\sigma>0$ is the scale parameter. 
The Lorenz curve and Gini coefficient associated with $LN(\mu,\sigma^2)$ are given by $L_{LN}(p|\sigma^2)=\Phi(\Phi^{-1}(p)-\sigma)$ and $G_{LN}=2\Phi(\sigma/\sqrt{2})-1$ where $\Phi(\cdot)$ and $\Phi^{-1}(\cdot)$ are the cumulative distribution function and quantile function of $N(0,1)$. 
It is well known that the Lorenz curve and Gini coefficient only depend on $\sigma$. 
For the latent process, it is simply set $u_t=\log(\sigma_t)$. 

\subsection*{Singh-Maddala distribution}
The probability density function of the Singh-Maddala distribution \citep{SM76} denoted by $SM(\alpha,\beta,\gamma)$ is given by 
\begin{equation*}
h_{SM}(x|\alpha,\beta,\gamma)=\frac{\alpha\gamma x^{\alpha-1}}{\beta^\alpha( 1+(x/\beta)^\alpha)^{\gamma+1}}, \quad x>0,
\end{equation*}
where $\alpha, \gamma>0$ are the shape parameters and $\beta>0$ is the scale parameter. 
The Lorenz curve and Gini coefficient are given by $L_{SM}(p|\alpha,\gamma)=\frac{B_{z}(1+1/\alpha,\gamma-1/\alpha)}{B(1+1/\alpha,\gamma-1/\alpha)}$ and $G_{SM}=1-\frac{\Gamma(\gamma)\Gamma(2\gamma-1/\alpha)}{\Gamma(\gamma-1/\alpha)\Gamma(2\gamma)}$ 
 where $z=1-(1-p)^{1/\gamma}$ and $B_t(a,b)=\int_0^t x^{a-1}(1-x)^{b-1}\d x$ is the incomplete beta function, $B(\cdot,\cdot)$ is the beta function and $\Gamma(\cdot)$ is the gamma function. 
The Lorenz curve and Gini coefficient do not depend on $\beta$. 
For the latent process, we set $\vu_t=(\log\alpha_t,\log\gamma_t)'$.

\subsection*{Dagum distribution}
The probability density function of the Dagum distribution \citep{Da77} denoted by $DA(\alpha,\beta,\kappa)$ is given by
\begin{equation*}
h_{DA}(x|\alpha,\beta,\kappa)=\frac{\alpha\kappa x^{\alpha\kappa-1}}{\beta^{\alpha\kappa}(1+(x/\beta)^\alpha)^{\kappa+1}}, \quad x>0,
\end{equation*}
where $\alpha,\kappa>0$ are the shape parameters and $\beta>0$ is the scale parameter. 
The Dagum distribution is also know to fit to income data well. 
The Dagum and Singh-Maddala distributions are the special cases of the mode flexible family called the generalised beta distribution of the second kind (GB2), denoted by $GB2(\alpha,\beta,\gamma,\kappa)$, and $DA(\alpha,\beta,\kappa)=GB2(\alpha,\beta,1,\kappa)$ and $SM(\alpha,\beta,\gamma)=GB2(\alpha,\beta,\gamma,1)$ . 
As for the Singh-Maddala distribution,  the Lorenz curve and Gini coefficient of the Dagum distribution do not depend on $\beta$ and are given by $L_{DA}(p|\alpha,\kappa)=\frac{B_z(\kappa+1/\alpha,1-1/\alpha)}{B(\kappa+1/\alpha,1-1/\alpha)}$ and $G_{DA}=\frac{\Gamma(\kappa)\Gamma(2\kappa+1/\alpha)}{\Gamma(\kappa+1/\alpha)\Gamma(2\kappa)}$. 
For the latent process, we set $\vu_t=(\log\alpha_t,\log\kappa_t)'$. 

\subsection*{Kakwani Lorenz curve}
The three-parameter Lorenz curve proposed by \citet{Ka80} is given by
\begin{equation*}
L_{KA}(p|\nu,\phi,\xi) = p - \nu p^\xi(1-p)^\delta,
\end{equation*}
where $\nu>0$ and $\xi,\delta\in(0,1]$. 
This is also called the beta type Lorenz curve and is known to be the one of the most flexible and best fitting Lorenz curves. 
The Gini coefficient is computed using the numerical integration based on \eqref{eqn:gini}. 
For the latent process, it is set $\vu_t=(\log\alpha_t, \log\frac{\xi_t}{1-\xi_t}, \log\frac{\delta_t}{1-\delta_t})'$.

\subsection*{Ortega Lorenz curve}
The Lorenz curve of \citet{OFLG91} is given by
\begin{equation*}
L_{OR}(p|\alpha,\delta)=p^\alpha(1-(1-p)^\delta),
\end{equation*}
where $\alpha\geq0$ and $\delta\in(0,1]$. 
When $\alpha=1$, this is the same as the Kakwani Lorenz curve with $\nu=\xi=1$. 
The Gini coefficient is computed using the numerical integration. 
For the latent process, it is set $\vu_t=(\log\alpha_t, \log\frac{\delta_t}{1-\delta_t})'$. 

\subsection*{Rasche Lorenz curve}
The Lorenz curve of \citet{RGKO80} is given by
\begin{equation*}
L_{RA}(p|\delta,\gamma)=(1-(1-p)^\delta)^\gamma,
\end{equation*}
where $\gamma>0$ and $\delta\in(0,1]$. 
The Gini coefficient is computed using the numerical integration.  
For the latent process, it is set $\vu_t=(\log\gamma_t, \log\frac{\delta_t}{1-\delta_t})'$. 


\end{document}